\documentclass[aps,preprint,preprintnumbers,amsmath,amssymb,nofootinbib]{revtex4}
\usepackage{amssymb}
\usepackage{lscape}
\usepackage{amsmath}
\usepackage{rotating}
\usepackage{bm}
\usepackage{graphicx}
\graphicspath{ {images/} }
\begin{document}
\title{Absorption of charged particles in a Reissner-Nordstr\"om Black-Hole: entropy evolution from Relativistic Quantum Geometry}
\author{$^{2}$ Marcos R. A. Arcod\'{\i}a\footnote{E-mail address: marcodia@mdp.edu.ar},  $^{2}$  Luis Santiago Ridao\footnote{E-mail address: santiagoridao@hotmail.com} and $^{1,2}$ Mauricio Bellini
\footnote{E-mail address: mbellini@mdp.edu.ar} }
\address{$^1$ Departamento de F\'isica, Facultad de Ciencias Exactas y
Naturales, Universidad Nacional de Mar del Plata, Funes 3350, C.P.
7600, Mar del Plata, Argentina.\\
$^2$ Instituto de Investigaciones F\'{\i}sicas de Mar del Plata (IFIMAR), \\
Consejo Nacional de Investigaciones Cient\'ificas y T\'ecnicas
(CONICET), Mar del Plata, Argentina.}
\begin{abstract}
Using Relativistic Quantum Geometry we show that the entropy can decrease in very small BHs, under certain circumstances, but always increases in very massive Black-Holes.
\end{abstract}
\maketitle

\section{Introduction and Motivation}

In 1973 Bekenstein\cite{B} suggested that the area ${\cal A}$ of the event horizon of a black hole is a measure of its entropy. This was a great advance in theoretical physical which was bolstered by Hawking's application of quantum field theory to black holes in 1975\cite{H}, when he deduced that these objects emit thermal radiation with a characteristic temperature has an entropy $S_{BH}$ determined by
\begin{equation}\label{ent}
S_{BH} = \frac{{\cal A}}{4 L^2_P},
\end{equation}
$L_P$ being the Planck length. In 2003, Davis {\em et al}\cite{da} studied the evolution of the cosmological horizon in some cosmological models. They showed that sometimes an apparent entropy decrease when black holes disappear over the cosmological event horizon. The aim of this work is to study the change of the entropy of a black hole when it is surrounded by matter, and absorb this matter. We shall
study the particular case where the black hole is charged, and matter surrounded the black hole has mass and also charge. To make it we shall work using Relativistic Quantum Geometry (RQG)\cite{rb}, with the aim to describe the variation of the entropy using a geometrical displacement from a Riemann to a Weyl manifold\cite{weyl}. In order to the Einstein tensor (and the Einstein equations), can be represented on
a Weylian manifold, the variation of the metric tensor must be done in a Weylian integrable manifold using an
auxiliary geometrical scalar field $\theta$. In the sense of the Riemannian geometry, the covariant derivative of the metric tensor is null, so that $\Delta g_{\alpha\beta}=g_{\alpha\beta;\gamma} \,dx^{\gamma}=0$\footnote{We shall denote with a $;$ covariant derivatives on a Riemann manifold and with a $|$ covariant derivatives on a Weyl manifold.}.

\section{Overview of RQG}

In this work
we shall consider a Weylian geometry characterized connections
\begin{equation}\label{ga}
\Gamma^{\alpha}_{\beta\gamma} = \left\{ \begin{array}{cc}  \alpha \, \\ \beta \, \gamma  \end{array} \right\}+ g_{\beta\gamma} \theta^{\alpha},
\end{equation}
such that the Weylian covariant derivative of the metric tensor is nonzero (non-zero non-metricity). For this reason [see the work \cite*{rb}] the Weylian variation of the metric tensor will be also nonzero\footnote{In what follows we shall denote with a $\Delta$ variations on the Riemann manifold, and with a $\delta$ variations on a Weylian manifold.}
\begin{equation}\label{gab}
\delta g_{\alpha\beta} = g_{\alpha\beta|\gamma} \,dx^{\gamma} = -\left[\theta_{\beta} g_{\alpha\gamma} +\theta_{\alpha} g_{\beta\gamma}
\right]\,dx^{\gamma},
\end{equation}
where
\footnote{We can define the operator
\begin{displaymath}
\check{x}^{\alpha}(t,\vec{x}) = \frac{1}{(2\pi)^{3/2}} \int d^3 k \, \check{e}^{\alpha} \left[ b_k \, \check{x}_k(t,\vec{x}) + b^{\dagger}_k \, \check{x}^*_k(t,\vec{x})\right],
\end{displaymath}
such that $b^{\dagger}_k$ and $b_k$ are the creation and destruction operators of space-time, such that $\left< B \left| \left[b_k,b^{\dagger}_{k'}\right]\right| B  \right> = \delta^{(3)}(\vec{k}-\vec{k'})$ and $\check{e}^{\alpha}=\epsilon^{\alpha}_{\,\,\,\,\beta\gamma\delta} \check{e}^{\beta} \check{e}^{\gamma}\check{e}^{\delta}$. }
\begin{equation}
dx^{\alpha} \left. | B \right> =  \hat{U}^{\alpha} dS \left. | B \right>= \delta\check{x}^{\alpha} (x^{\beta}) \left. | B \right> ,
\end{equation}
is the eigenvalue that results when we apply the operator $ \delta\check{x}^{\alpha} (x^{\beta}) $ on a background quantum state $ \left. | B \right> $, defined on the Riemannian manifold\footnote{In our case the background quantum state can be represented in a ordinary Fock space in contrast with LQG, where operators is qualitatively different
from the standard quantization of gauge fields.}. Furthermore, $\hat{U}^{\alpha}$ are the components of the Riemannian
velocities. Here, we denote with a {\it hat} the quantities represented on the Riemannian background manifold.  The Weylian line element is given by
\begin{equation}
dS^2 \, \delta_{BB'}=\left( \hat{U}_{\alpha} \hat{U}^{\alpha}\right) dS^2\, \delta_{BB'} = \left< B \left|  \delta\check{x}_{\alpha} \delta\check{x}^{\alpha}\right| B'  \right>.
\end{equation}
Hence, the differential Weylian line element $dS$ provides the displacement of the quantum trajectories with respect to the classical (Riemannian) ones. When we displace
with parallelism some vector $v^{\alpha}$ on the Weylian manifold, we obtain
\begin{equation}
\delta v^{\alpha} = \theta^{\alpha} g_{\beta\gamma} v^{\beta} dx^{\gamma}, \qquad \rightarrow \qquad \frac{\delta v^{\alpha}}{\delta S} =
\theta^{\alpha} v^{\beta}\,g_{\beta\gamma}\, \hat{U}^{\gamma},
\end{equation}
where we have taken into account that the variation of $v^{\alpha}$ on the Riemannian manifold, is zero: $\Delta v^{\alpha}=0$. From the action's point of view, the scalar field $\theta(x^{\alpha})$ drives a geometrical displacement from a Riemannian manifold to a Weylian one, that leaves the action invariant
\begin{eqnarray}
{\cal I} &=& \int d^4 x\, \sqrt{-\hat{g}}\, \left[\frac{\hat{R}}{2\kappa} + \hat{{\cal L}}\right] \nonumber \\
&=& \int d^4 x\, \left[\sqrt{-\hat{g}} e^{-2\theta}\right]\,
\left\{\left[\frac{\hat{R}}{2\kappa} + \hat{{\cal L}}\right]\,e^{2\theta}\right\}, \label{aact}
\end{eqnarray}
where $\hat{R}$ is the Riemannian scalar curvature, $\kappa= 8\pi G$, $G$ is the gravitational constant and $\hat{\cal L}$ is the matter lagrangian density on the background Riemmanian
manifold. If we require that $\delta {\cal I} =0$, we obtain
\begin{equation}
-\frac{\delta V}{V} = \frac{\delta \left[\frac{\hat{R}}{2\kappa} + \hat{{\cal L}}\right]}{\left[\frac{\hat{R}}{2\kappa} + \hat{{\cal L}}\right]}
= 2 \,\delta\theta,
\end{equation}
where $\delta\theta = \theta_{\mu} dx^{\mu}$ is an exact differential and $\hat{ V}=\sqrt{-\hat{ g}}$ is the volume of the Riemannian manifold. Of course, all the variations are in the Weylian geometrical representation, and assure us gauge invariance because $\delta {\cal I} =0$. This means that the Weylian volume in the second row of the side of (\ref{aact}) will be
$V=\hat{V} \, e^{-2\theta}$.

\section{Relativistic dynamics on a Weylian manifold}

The Einstein tensor can be written as\cite*{rb}
\begin{equation}
\bar{G}_{\alpha\beta} = \hat{G}_{\mu\nu} + \theta_{\alpha ; \beta} + \theta_{\alpha} \theta_{\beta} + \frac{1}{2} \,g_{\alpha\beta}
\left[ \left(\theta^{\mu}\right)_{;\mu} + \theta_{\mu} \theta^{\mu} \right],
\end{equation}
where we have made use of the fact that the connections are symmetric. Here, $\bar{G}_{\mu\nu}$ are the Weylian components of the Einstein tensor and $\hat{G}_{\mu\nu}$ the Riemannian (background) ones, such that $\bar{G}_{\alpha\beta} = \hat{G}_{\mu\nu}-\hat{g}_{\mu\nu} \Lambda$, where $\Lambda$ is the cosmological constant. As can be demonstrated: $ \left(\theta^{\mu}\right)_{;\mu}\equiv \hat\Box \theta=0$ and $\theta_{\mu} \theta^{\mu}=-(4/3) \Lambda $. Notice that $\Lambda$ is a Riemannian invariant but not a Weylian one: $\Lambda\equiv \Lambda(\theta, \theta_{\alpha})$. Hence, we can consider a functional
\begin{equation}\label{aa}
\Lambda(\theta, \theta_{\alpha}) = -\frac{3}{4} \left[ \theta_{\alpha} \theta^{\alpha} + \hat{\Box} \theta\right],
\end{equation}
to define a geometrical quantum action on the Weylian manifold
\begin{equation}
{\cal W} = \int d^4 x \, \sqrt{-g} \,\, \Lambda(\theta, \theta_{\alpha}).
\end{equation}
The dynamics of the geometrical field $\theta $, is given by the Euler-Lagrange equations, after imposing $\delta
{\cal W}=0$:
\begin{equation}
\frac{\delta \Lambda}{\delta \theta} - \hat{\nabla}_{\alpha} \left( \frac{\delta \Lambda}{\delta \theta_{\alpha}}\right) =0,
\end{equation}
where the variations are defined on the Weylian manifold. One of the interesting consequences of this fact is the explanation of the variation of the cosmological constant $\Lambda$ along the evolution of the universe. For instance, if we take as the initial volume of the primordial universe the Planckian volume: $\hat{V} = V_P$, we obtain that the present day volume $V=\frac{4}{3} \pi \left(\frac{c}{H_0}\right)^3$ is $10^{183}$ times bigger than $\hat{V}$, where we have taken $\frac{c}{H_0} = 1.27 \times 10^{26} \,{\rm meters}$. This expression can be rewritten as
\begin{equation}
V=\hat{V} \, e^{\frac{\Lambda S^2}{3}} ,
\end{equation}
where the value of the present day cosmological constant, is $\Lambda=\frac{3 H^2_0}{c^2} = 1.8621 \times 10^{-52} \, {\rm meters}^{-2}$. Hence, if we use the fact that $\delta
{\cal W}=0$, we obtain that $\frac{\delta\Lambda}{{\Lambda}}=-\frac{\delta V}{V}$, so that the value of the cosmological constant at the Planck time, should must been
\begin{equation}
\Lambda_P = 1.8621 \times 10^{131} \,{\rm meters}^{-2}.
\end{equation}
This means that $\delta\Lambda\neq 0$, but $\Delta\Lambda=0$.

Furthermore, $ \Pi^{\alpha}=\frac{\delta \Lambda}{\delta \theta_{\alpha}}=-\frac{3}{4} \theta^{\alpha}$ is the geometrical momentum and the
dynamics of $\theta$ describes a free scalar field
\begin{equation}\label{si}
\hat{\Box} \theta =0,
\end{equation}
so that the momentum components $\Pi^{\alpha}$ comply with
the equation
\begin{equation}
\hat{\nabla}_{\alpha} \Pi^{\alpha} =0.
\end{equation}
If we define the scalar invariant
$\Pi^2=\Pi_{\alpha}\Pi^{\alpha}$, we obtain that
\begin{equation}
\left[\theta,\Pi^{2}\right] = \frac{9}{16}\left\{ \theta_{\alpha} \left[\theta,\theta^{\alpha} \right]
 + \left[\theta,\theta_{\alpha} \right] \theta^{\alpha} \right\}=0,
\end{equation}
where we have used that $\theta_{\alpha} U^{\alpha} = U_{\alpha} \theta^{\alpha}$, and
\begin{eqnarray}\label{con}
\left[\theta(x),\theta^{\alpha}(y) \right] &=& - i \Theta^{\alpha}\, \delta^{(4)} (x-y), \nonumber \\
\left[\theta(x),\theta_{\alpha}(y) \right] &=&
i \Theta_{\alpha}\, \delta^{(4)} (x-y),
\end{eqnarray}
with $\Theta^{\alpha} = \hbar\, \hat{U}^{\alpha}$. Therefore we can define
the relativistic invariant $\Theta^2 = \Theta_{\alpha}
\Theta^{\alpha} = \hbar^2 \hat{U}_{\alpha}\, \hat{U}^{\alpha}$.
Additionally, it is possible to define the Hamiltonian operator
\begin{equation}
{\cal H} = \left(\frac{\delta \Lambda}{\delta \theta_{\alpha}}\right) \theta_{\alpha} - \Lambda(\theta,\theta_{\alpha}),
\end{equation}
such that the eigenvalues of quantum energy becomes from ${\cal H}\left|B\right> = E\left|B\right>$. Can be demonstrated that
$\delta{\cal H}=0$, so that the quantum energy $E$ is a Weylian invariant. \\

\section{Entropy evolution of a Reissner-Nordstr\"om Black-Hole (RNBH) by absorption of charged particles}

We consider a RNBH with mass $M$ and squared
electric charge $Q^2$, such that outside the horizon radius $r_+=GM \left[1-\left(1-\left(\frac{2 Q}{G M}\right)^2\right)^{1/2}\right]$,
there are massive charged particles which are
absorbed by the BH. The line element is given by
\begin{equation}\label{bh}
dS^2 = f(r) dt^2 - \frac{1}{f(r)} dr^2 - r^2\, d\Omega^2,
\end{equation}
where $d\Omega^2 = \sin^2\theta\,d\phi^2 + d\theta^2$ is the
square differential of solid angle and $f(r) = 1-\frac{2G M}{r}
+ \frac{Q^2}{r^2}$, such that $(G M)^2 \geq (Q)^2$ and $Q=\frac{q}{4\pi \epsilon_0}$.

Making use of the fact that $\delta {\cal W}=0$, we can obtain the relationship between the scalar flux and the change of area $\delta {\cal A}$ of the black-hole
\begin{equation}\label{area}
\frac{d\theta}{d S} = -\frac{3}{S} {\rm ln}\left[ \frac{\delta{\cal A}}{\bar{{\cal A}}} +1\right],
\end{equation}
where $\frac{\delta{\cal A}}{{\bar{\cal A}}}= 2 \frac{r_+}{\bar{r}^2_+ } \delta r_+$, $\frac{\delta{\cal A}}{\bar{{\cal A}}} +1=\left(\frac{r_{+}}{\bar{r}_{+}}\right)^2 = e^{\frac{2}{9} \Lambda S^2} \geq 1$, and the square of the radius between the final $r_+$ and the initial horizon radius $\bar{r}_+$ is related with the change of area. We can relate the area of the horizon radius to the Bekenstein-Hawking entropy (\ref{ent}). Using the fact that $L_P=\sqrt{G \hbar/c^3}=1.616 \times 10^{-35}\,{\rm meters}$, the expression (\ref{area}) can be rewritten as
\begin{equation}
\frac{d\theta}{d S} = -\frac{3}{S} {\rm ln}\left[ \frac{\delta{\cal S}_{BH}}{\bar{{\cal S}}_{BH}} +1\right],
\end{equation}
or
\begin{equation}
 \frac{\delta{\cal S}_{BH}}{\bar{{\cal S}}_{BH}} = e^{\frac{2}{9} \Lambda S^2}-1.
\end{equation}
This result can be expressed in terms of the volume of the manifold evaluated at the horizon radius:
\begin{equation}\label{vol1}
 \frac{\delta{\cal S}_{BH}}{\bar{{\cal S}}_{BH}} = \left.\left(\frac{V}{\hat{V}}\right)^{2/3} - 1\right|_{r=r_+}.
\end{equation}
This is a relevant result because describes in general the change of the horizon entropy as a function of the radius between the final and initial volumes related to the horizon radius. It is obvious from (\ref{vol1}) that an increasing volume increases $S_{BH}$ and vice versa. Moreover, using the fact that $\delta{\cal W}=0$, we obtain that the change of volume of the Weylian manifold (the black-hole's increasing of volume evaluated at the horizon radius), with respect to the Riemannian one $\sqrt{-\hat{g}}$, is
\begin{equation}\label{vol}
V = \hat{V}\, e^{- \int d\theta} = \hat{V}\, e^{\frac{\Lambda S^2}{3}},
\end{equation}
where $\Lambda S^2 >0$. This means that $V \geq \sqrt{-\hat{g}}$, for $S^2\geq 0$, $\Lambda >0$ and $d\theta <0$. Therefore, we require a metric's signature  $(-,+,+,+)$ in order for the
cosmological constant to be positive and a metric's signature $(+,-,-,-)$ in order to have $\Lambda \leq 0$.

The variation of mass can be expressed in terms of the area and charge variations
\begin{equation}
\delta M = K \frac{\delta {\cal A}}{8\pi G} + \frac{Q \delta Q}{r_+},
\end{equation}
where $K$ is the temperature on the horizon radius.

When we study the entropy evolution $\frac{\delta {\cal A}}{{\cal A}}$, we distinguish two different cases
\begin{itemize}
\item When $\frac{d\theta}{dS} >0$, one obtains that $\frac{\delta {\cal A}}{{\cal A}}<0$, and must be fulfilled that
\begin{equation}
\delta M < \frac{Q \delta Q}{G M}.
\end{equation}
This condition is always fulfilled when the absorbed charge has the same sign than the inner charge of the BH. Therefore, if the charge of the BH is positive (negative), is necessary that only positive (negative) charge to be absorbed by the BH, in order for the entropy to be decreasing. This condition will be favored by very small BHs.\\

\item When $\frac{d\theta}{dS} <0$, one obtains that $\frac{\delta {\cal A}}{{\cal A}} >0$, and must be fulfilled that
\begin{equation}
\delta M > \frac{2 Q \delta Q}{G M}.
\end{equation}
This condition is always fulfilled for any sign of charges absorbed by the BH with any sign of charge in its interior, and will be favored by very massive BHs.
\end{itemize}

\section{Final comments}

In RQG the dynamics of a geometrical scalar field defined in a Weyl integrable manifold preserves the gauge-invariance under transformations of the Einstein
equations and the geometrical vector fields\cite{be}, that involves the cosmological constant. Using this formalism, we have demonstrated that [see equation (\ref{vol1})], an increasing volume of the horizon radius drives an increment of the Bekenstein-Hawking entropy, and vice versa. In particular, we have studied the evolution of a RNBH's entropy when it is surrounded by charged particles. In particular, if the charge of the BH is positive (negative), is necessary that only positive (negative) charge to be absorbed by the BH, in order for the entropy to be decreasing. The interesting result is that under certain circumstances this entropy can decrease. This behavior is more plausible in small BHs.
\vskip .5cm
\noindent
{\bf Acknowledgments}\\ \\
\noindent
The authors acknowledge UNMdP and CONICET for financial support.\\


\begin{thebibliography}{99}
\bibitem{B} J. D. Bekenstein, Phys. Rev. {\bf D7} (1973) 2333.
\bibitem{H} S. W. Hawking, Comm. Math. Phys. {\bf 43} (1975) 199.
\bibitem{da} T. M. Davis, P. C. W. Davies, Ch. H. Lineweaver, Class. Quant. Grav. {\bf 20} (2003) 2753.
\bibitem{rb} L. S. Ridao, M. Bellini, Phys. Lett. {\bf B751}: (2015) 565.
\bibitem{weyl} H. Weyl, Sitzungesber Deutsch. Wiss. Berlin, 465 (1918); \\
H. Weyl, {\em Space, Time, Matter} (Dover, New York, 1952).
\bibitem{be} M. Bellini, Phys. Dark Univ. {\bf 11} (2016) 64.
\end{thebibliography}
\end{document}